\documentclass[preprint,superscriptaddress,amsmath,amssymb,aps,physrev,prb,floatfix, fleqn]{revtex4-2} \usepackage{bm} 
\usepackage{hyperref}
\usepackage[english]{babel}
\usepackage[utf8]{inputenc}   
\usepackage{xspace}
\usepackage{tensor} 
\usepackage{array,color}
\usepackage{relsize,exscale}  
\usepackage{textcase}	
\usepackage{soul,color}		
\usepackage{calc} 
\usepackage{makeidx,showidx}
\usepackage{graphicx}
\usepackage{wrapfig}
\usepackage{textcomp}
\usepackage{siunitx}
\usepackage[version=3]{mhchem}
\usepackage{enumitem}

\newcommand{\lsco} {{La$_{2-x}$Sr$_x$CuO$_4$}\@\xspace}

\newcommand{\ybcoF} {$\ce{YBa2Cu3O_{7}}$\@\xspace}
\newcommand{\ybco} {$\ce{YBa2Cu3O_{6+y}}$\@\xspace}

\newcommand{\ybcoE} {$\ce{YBa2Cu4O8}$\@\xspace}

\newcommand{\tc} {\ensuremath{T_{\mathrm c}}\@\xspace}
\newcommand{\cperp}{\ensuremath{c \bot B_0}\@\xspace}
\newcommand{\cpara}{\ensuremath{{c\parallel\xspace B_0}}\@\xspace}
\newcommand{\tcmax} {\ensuremath{T_{\mathrm{c, max}}}\@\xspace}

\newcommand{\beq} {\begin{equation}}
\newcommand{\eeq} {\end{equation}}

\preprint{Lee NMR}
\begin{document}

\title{\textbf{Pseudogap and Condensation in Cuprate Superconductors from NMR Shifts}}
\author{Abigail Lee}
\author{Jürgen Haase}
\email{juergen.haase@uni-leipzig.de}
\affiliation{University of Leipzig, Felix Bloch Institute for Solid State Physics, Linn\'estr.\@ 5, 04103 Leipzig, Germany}

\date{\today}
\medskip

\begin{abstract}
\vspace{1cm}
{The electronic properties of the high-temperature superconducting cuprates are encoded in complex sets of NMR data, but without microscopic theory, reliable NMR phenomenologies are in demand. Early analyses of NMR could only focus on very few materials and discovered spin singlet pairing and the enigmatic pseudogap. However, a coherent phenomenology of shift and relaxation could not be established, as incoming data from other cuprates complicated the picture. Today, due to work of many groups worldwide, planar copper and oxygen NMR data are available for most cuprates. Here, based only on symmetry of the two Cu hyperfine couplings, an anisotropic $A_\alpha$ and isotropic $B$, the Cu shifts are disentangled, and two different shift components emerge. Upon doping the cuprates, metallic B-spins are created above the pseudogap temperature which is shared with metallic A-spins. Further doping decreases the pseudogap temperature and increases the B-spin, but less so the A-spin. The apparent linear rate of increase in density of states of the B-spin with doping increases nearly threefold above about $x=0.20$, where the pseudogap has disappeared and A and B turn into superconducting metals, i.e. they disappear rapidly at $T_\mathrm{c}$. The pseudogap temperature is a measure of the coupling between A and B, which suppresses the shifts but not nuclear relaxation. Spin singlet pairing involves A and B according to three simple rules for condensation which will be discussed. The optimal $T_\mathrm{c}$ demands a special match between A and B and involves systems with a pseudogap. However, the highest $T_\mathrm{c}$ of all cuprates is not encoded in the shift, but rather in nuclear relaxation and charge sharing between planar Cu and O. Relations to other probes are discussed.}
\end{abstract}

{\flushleft PACS numbers:  74.72.-h, 74.25.Jb, 74.25.-q, 74.25.Nf}\\

\maketitle

\newpage

\section{Introduction}
Electronic phases of a material leave detailed fingerprints in nuclear spin quantum states. For example, changes in energies and lifetimes \cite{Slichter1990} cause in metals a temperature ($T$) independent NMR Knight shift ($K$) in a magnetic field $B_0$ (spin polarization from Pauli spin susceptibility), which is related to nuclear relaxation ($1/T_1$) by the Korringa law:  $1/T_1 T = (\gamma_n/\gamma_e)^2 4\pi k_\mathrm{B}/\hbar \cdot K^2$ \cite{Korringa1950}. For conventional superconductors, there is the Hebel-Slichter coherence peak in relaxation \cite{Hebel1957} just below the critical temperature of superconductivity ($T_\mathrm{c}$), and the ensuing drop in relaxation and shift (Yosida function \cite{Yosida1958}) as the temperature drops further below \tc (spin singlet pairing). However, without microscopic theory, as for the high-temperature superconducting cuprates \cite{Bednorz1986}, it is difficult to interpret NMR, in particular in the rather complex cuprates. 

\begin{figure}
\centering
\includegraphics[width=.5\textwidth ]{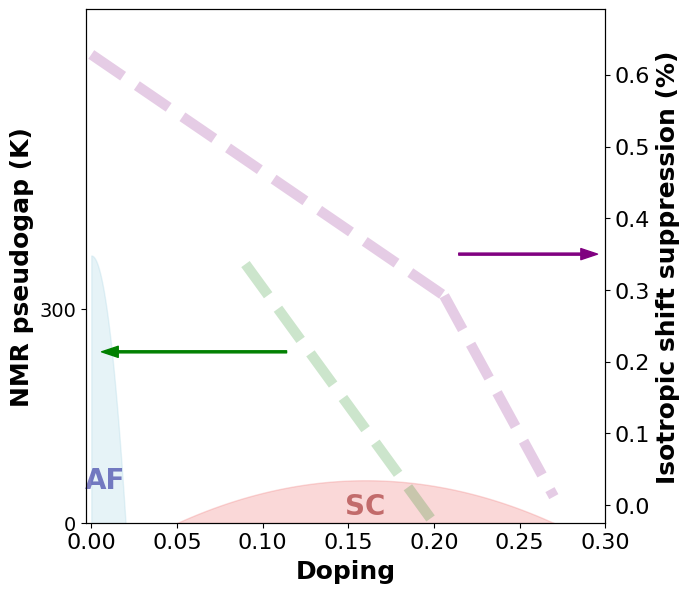}
\caption{The cuprate normal state properties, outside the antiferromagnetic (AF) and superconducting (SC) phases in the phase diagram, are still under debate. Here the planar Cu isotropic shift component (purple dashed line) is shown to traverse the phase diagram as a function of doping, in addition to the NMR pseudogap (green dashed line). The former likely just measures density of states of a metallic component as a function of doping, but may also enable coupling to another component to create the pseudogap phenomenon (the NMR pseudogap line, $\Delta_{\mathrm{NMR}}/k_B$, is estimated for a V-shaped, temperature independent gap). The two phenomena have to be distinguished as they also influence condensation (more details are given in the main text).}\label{fig:fig1}
\end{figure}

Early cuprate NMR shift and relaxation studies of a few \ybco samples played a decisive role in setting up fundamental behavior through $^{63,65}$Cu \cite{Walstedt1988,Mali1987} and $^{17}$O and $^{89}$Y \cite{Alloul1989} NMR. For example, planar O and Y show a superconducting metal for \ybcoF (just beyond optimal doping at $y=0.96$). This means that above \tc the shifts are temperature independent, and the relaxation is proportional to temperature, connected by the Korringa relation. A missing Hebel-Slichter peak just below \tc points to unconventional superconductivity. Shift and relaxation disappear at the lowest temperatures, signaling spin singlet pairing. 

Luckily, NMR investigations continued into many more cuprates. The new data are at odds with the old phenomenology. It became clear, for example, that a single, temperature dependent electronic spin to which all nuclei couple cannot explain the data in the previously assumed way \cite{Slichter2007,Haase2009b,Haase2012,Rybicki2015}. Today, we have data for more than 3 dozen materials, which ultimately enables us to uncover rather different, universal phenomenology that we find substantially different from previous accounts, as our recent discussion of nuclear relaxation already shows \cite{Lee2026}.
 
The NMR shifts discussed here have mostly been gathered and presented before, for planar Cu \cite{Haase2017, Avramovska2019,Avramovska2022} and planar O \cite{Nachtigal2020,Avramovska2022}. Essentially they show that the planar Cu shifts do not even scale for different directions of the field, nor with planar O. Very recently, scaling between planar Cu axial and all O shifts was uncovered, clearly demonstrating the presence of a second isotropic shift component at planar Cu \cite{Bandur2026}. However, it was based on common assumptions for the size of the hyperfine constants, which will be relaxed in the analysis shown here. 

Here we build on two atomic hyperfine coefficients for Cu, each relating to an independent electronic spin, without making special assumptions about their size. As a consequence, we find a doping dependent, isotropically coupled metal and a rather doping independent (more family-dependent), anisotropically coupled 2nd metallic component. These two components are characterized across the cuprates and how they relate to the pseudogap and condensation, as well as to the nuclear relaxation, which has already been addressed \cite{Lee2026}. This all gives a very robust NMR phenomenology of the cuprates, mostly in metallic terms which also work rather well to describe nuclear relaxation \cite{Lee2026}, and should help interpret other results and advance theory, which could be much more involved.

\section{Decomposition of the shifts}
Most of the Cu NMR shifts, ${^{63}K}_{\parallel, \perp}$ with the magnetic field parallel (\cpara) and perpendicular (\cperp) to the crystal $c$-axis respectively, from the CuO$_2$ plane have been presented and discussed recently \cite{Haase2017,Nachtigal2020,Avramovska2022,Bandur2026}. 
Here we just use a different decomposition of the shifts based on the two atomic hyperfine coefficents' symmetry. First, there is the well-known anisotropic, onsite hyperfine coupling of the Cu$^{2+}$ ion, $A_\alpha$, from spin associated with the $3d(x^2-y^2)$ orbital (dipolar, core polarization and spin-orbit terms), with $A_\parallel$ being negative and of known anisotropy. That is the reason why one expected a negative ${^{63}K}_{\parallel}$ which was not found \cite{Slichter2007,Walstedt2008}, but which we will identify here. We use an anisotropy of $\left|A_\parallel\right|/A_\perp = 6$ \cite{Mila1989b,Takigawa1989b,Pennington1989,Renold2003}, but the exact value is not essential for the overall shift phenomenology. Second, there is the isotropic hyperfine coupling, $B$, based on Cu 4s spin density \cite{Mila1989b,Pavarini2001}, with no assumptions about transfer of spin density. 
In the decomposition, we allow, but don't require, the spin components related to the two hyperfine coefficients, $\chi_\mathrm{A}(T)$ and $\chi_\mathrm{B}(T)$, to be independent. That is, we use the these shift equations to disentangle all Cu shifts,
\begin{align}
{^{63}K}_\perp (T) & = B \chi_{\rm B}(T)+ A_\perp \chi_{\rm A}(T)  \label{eq:s2}\\
{^{63}K}_\parallel (T) &= B \chi_{\rm B}(T) + A_\parallel \chi_{\rm A}(T)\label{eq:s3}.
\end{align}
The isotropic term is present in both Cu shifts, and the anisotropic term with its negative sign will lead to a decrease of ${^{63}K}_{\parallel}$ compared to ${^{63}K}_{\perp}$. This explains the complicated Cu shifts that are shown in Fig.~\ref{fig:fig2}(c, d) together with sketched results to help the overall understanding (a full account is given in the Appendix). 

A first, very important finding is the following: At high doping levels, there are metallic high-temperature plateaus of $B\chi_\mathrm{B}$ and $A_\alpha\chi_\mathrm{A}$, as if each originates from a metal. However, only $B\chi_\mathrm{B}$ decreases significantly with lowering doping (not family) beginning near the entrance to the superconducting dome, depicted in Fig.~\ref{fig:fig1}. $A_\alpha \chi_\mathrm{A}$ changes little. It then appears that the density of states (DOS) of the $\chi_\mathrm{B}$ metal drops until about optimal doping, while its overall temperature dependence remains the same (that of a metal with different DOS). Below about $x\approx0.2$ the high temperature metallic plateau of $B\chi_\mathrm{B}$ remains clearly visible even at much lower doping levels, but it drops with a much lower rate as function of doping ($\Delta_x \chi_\mathrm{B}(300\text{K}) / \Delta x \approx 0.61$ below and about 1.65 above optimal doping).

\begin{figure}[h]
\centering
\includegraphics[width=.8\textwidth]{fig2.png}
\caption{Two Cu shift components, from two hyperfine coefficients. (a,b) Sketch of how the isotropic and anisotropic shift components, $B\chi_\mathrm{B}(T)$ and $|A_\parallel| \chi_\mathrm{A}(T)$, evolve with doping (the sign of the negative $A_\parallel$ has been removed for easier visual comparison). Sufficiently overdoped materials show a temperature independent shift above $T_\mathrm{c}$ (marked with arrows), as expected from a metal. The isotropic component continues to increase beyond about optimal doping. At lower doping, the pseudogap broadens both shifts near $T_\mathrm{c}$, rendering both temperature dependent above $T_\mathrm{c}$, increasingly so as doping decreases. Note that, as the temperature is lowered, the anisotropic shift component can vanish while the isotropic shift is still decreasing. (c,d) Experimental Cu shifts (See Appendix for data references) measured with the magnetic field perpendicular (\cperp) and parallel (\cpara) to the crystal $c$-axis, respectively. The shift for \cperp is dominated by (a). The shift for \cpara is the difference between both, cf. \eqref{eq:s2} and \eqref{eq:s3}. }\label{fig:fig2}
\end{figure}

Note that for \ybcoF, a heavily investigated material and cornerstone of early NMR,  the planar Cu isotropic shift (not that of planar O) is already strongly suppressed. 

A second, very important observation is the following: It is well-known that strongly doped cuprates show a sudden drop in shift at \tc, expected from spin singlet pairing. However, this behavior begins to disappear below optimal doping, cf.~Fig.~\ref{fig:fig2}. In fact, with decreasing doping, a broadening at \tc sets in and increasingly smears out the sudden drop for both spin components. Eventually the broadening extends to far above (and below) \tc. This leads to a temperature dependence above \tc. Now, in view of the doping induced changes of the high-temperature plateau of $B\chi_\mathrm{B}$, we argue that this plateau keeps its physical origin (being a metal) also at  lower doping levels, i.e.\@ as long as the plateau is still visible. It is the broadening at \tc that creates an apparent decrease of the shift above \tc. This is the pseudogap effect in NMR shifts. Note that it is absent in the Cu relaxation data in the sense that they always show a rather sharp transition at \tc, which appears out of a metallic temperature dependence in a doping-dependent temperature range above \tc \cite{Lee2026}. 

\begin{figure}[h]
\centering
\includegraphics[width=0.8\textwidth ]{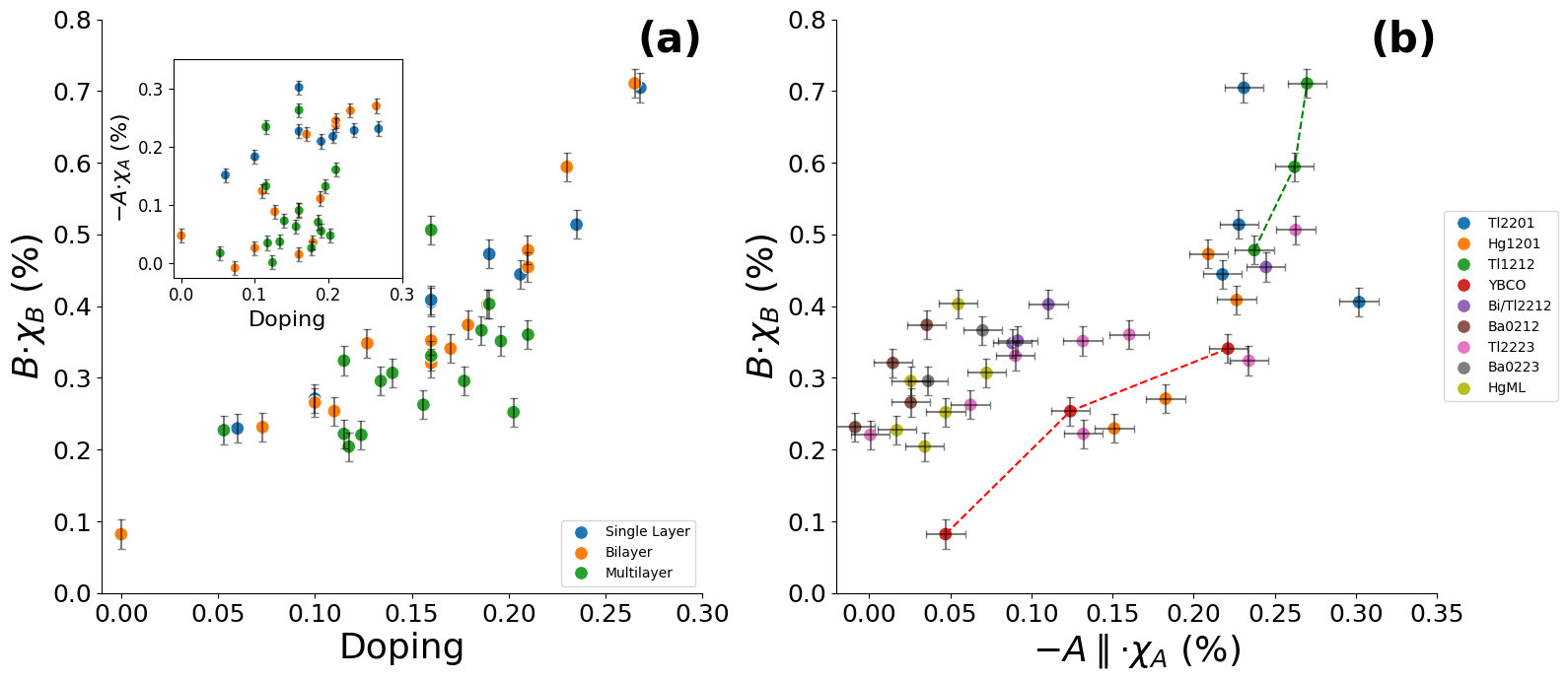}
\caption{(a) High-temperature $B\chi_{\mathrm{B}}$ for single-, bi-, and multilayer materials plotted against doping. Inset: High-temperature $-A_{\parallel}\chi_{\mathrm{A}}$ plotted against doping (note the lack of clear doping dependence). (b) High-temperature $B\chi_{\mathrm{B}}$ plotted against high-temperature $-A_{\parallel}\chi_{\mathrm{A}}$ for a number of different families. $B\chi_{\mathrm{B}}$ is never smaller than $-A_{\parallel}\chi_{\mathrm{A}}$. A few families have been connected with dashed lines as guide to the eye to show their different trajectories. Note the scale of the axis for $-A_{\parallel}\chi_{\mathrm{A}}$ is not the same as that for $B\chi_{\mathrm{B}}$.}\label{fig:fig3}
\end{figure}

The decomposition of the Cu temperature dependent shift employed here leaves a small uncertainty with respect to the size of $\chi_\mathrm{A}$. Nevertheless, it appears that a slight family dependence of this component is real, cf~Fig.~\ref{fig:fig3} (note also, that these are the literature data of all cuprates from around the world, adjusted for a common shift reference \cite{Haase2017}). 

Despite the very different magnitude of both components at high temperatures, they assume strict temperature dependencies, which is seen by plotting $|A_\parallel|\chi_\mathrm{A}(T)$ vs $B\chi_\mathrm{B}(T)$, with $T$ as implicit parameter. Whenever the shifts become temperature dependent, they adopt one of three slopes. These slopes have been noted before and addressed in terms of the anisotropic shifts \cite{Haase2017}. In our decomposition, the three slopes transform into, 
\begin{align}
\Delta_T\chi_\mathrm{A}(T) &\approx0,\;\;  \Delta_T\chi_\mathrm{B}(T) \neq0\label{eq:slope1}\\
\Delta_T\chi_\mathrm{A}(T) &\approx\frac{B}{|A_\parallel |}\cdot\Delta_T\chi_\mathrm{B}(T)\label{eq:slope2}\\
\Delta_T\chi_\mathrm{A}(T) &\approx0.5\frac{B}{|A_\parallel |}\cdot\Delta_T\chi_\mathrm{B}(T).\label{eq:slope3}
\end{align}
Note that these three slopes are simply the transformation of those of $^{63}K_\perp(T)$ vs $^{63}K_\parallel(T)$ under this decomposition. We find the temperature dependence of planar O shift to be proportional to $\chi_\mathrm{A}$ as shown before \cite{Bandur2026}. Since the planar O shift is known to have a doping independent anisotropy, we can use the most commonly measured shift with the magnetic field parallel to the crystal $c$-axis \cite{Avramovska2022b}, and write,
\begin{equation}
{^{17}K}_{\rm c} (T)  = C \chi_\mathrm{A}(T).\label{eq:s4}
\end{equation}

Another salient observation from our decomposition is that $-A_\parallel\chi_A(T)$ is never larger than $B\chi_B(T)$ for any of the materials which have been examined here. In fact, \ybcoE nearly traces out this limiting condition, as does \lsco (not shown in the figure, as it follows the same line as \ybcoE, but requires the subtraction of an additional shift contribution to $^{63}K_{\perp}$, which adds uncertainty).

As was previously noted \cite{Haase2017, Bandur2026}, the optimally-doped and higher-\tc materials tend to follow a slope of 5/2 in  $^{63}K_\perp(T)$ vs $^{63}K_\parallel(T)$, which in this framework corresponds to the ratio $\Delta_T\chi_\mathrm{A}(T) \approx0.5\frac{B}{|A_\parallel |}\Delta_T\chi_\mathrm{B}(T)$. The highest \tc materials follow this ratio of temperature dependencies for the entire temperature-dependent regime. Underdoped materials tend to follow, at least for some temperature range, $\Delta_T\chi_\mathrm{A}(T)\approx\frac{B}{|A_\parallel |}\Delta_T\chi_\mathrm{B}(T)$, while overdoped materials, which have far larger high temperature plateau in $B\chi_\mathrm{B}$ than $|A|\chi_\mathrm{A}$, typically show temperature ranges where only $\chi_\mathrm{B}(T)$ changes. This is similar to the matching condition and its relation to the superconducting dome and \tcmax which was described in \cite{Bandur2026}. 

\begin{figure}[h]
\centering
\includegraphics[width=.7\textwidth ]{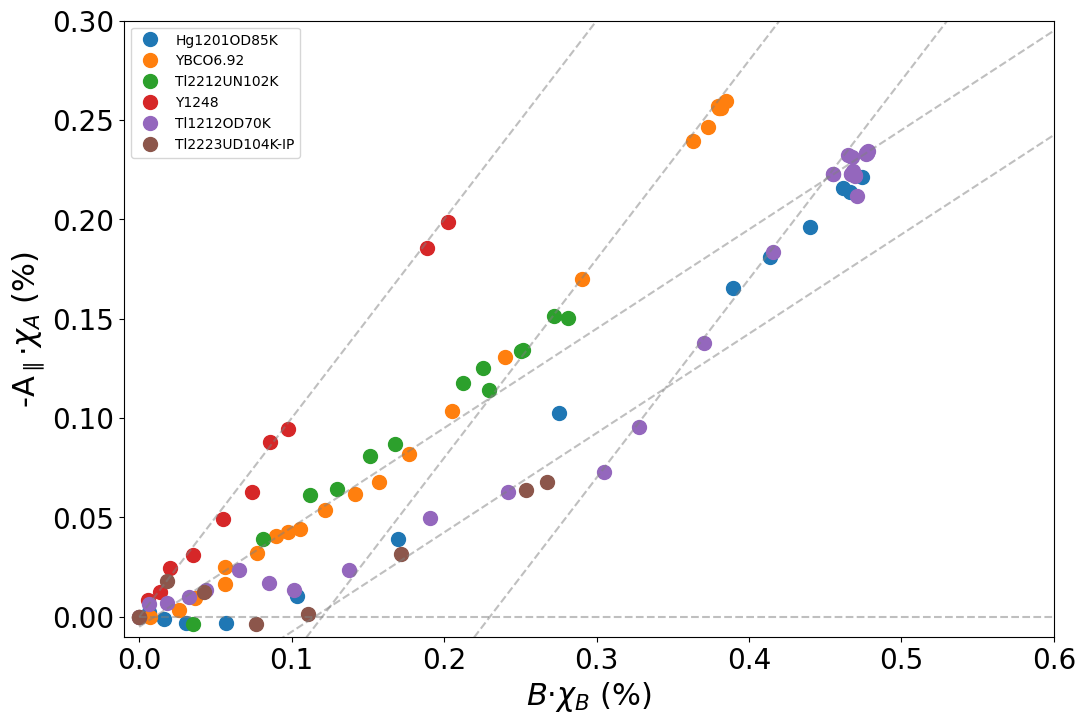}
\caption{Temperature-implicit plot of $-A_\parallel\chi_\mathrm{A}(T)$ vs $B\chi_\mathrm{B}(T)$. Dashed lines are guides to the eye to indicate the typical slopes according to \eqref{eq:slope1} to \eqref{eq:slope3}, see main text for details.}\label{fig:fig4}
\end{figure}

\section{Discussion}
{\flushleft The decomposition} of the shifts in terms of the two essential hyperfine coefficients results in two independent spin components from $\chi_\mathrm{A}(T)$ and $\chi_\mathrm{B}(T)$. The former gives the negative spin shift on ${^{63}K}_\parallel$ and a small contribution to ${^{63}K}_\perp$, while $\chi_\mathrm{B}$ contributes as isotropic term to both shifts equally. $\chi_\mathrm{A}$ is also responsible for the planar O shift \cite{Bandur2026}. 

At high doping levels, both components are nearly indistinguishable from superconducting metals (the B-component is significantly larger than the A-component), i.e., both show a high-temperature plateau above \tc and rapidly disappear below \tc, as expected for spin singlet pairing. Near the beginning of the superconducting dome, the B-component metal plateau rapidly decreases as if the DOS is falling  with decreasing doping. The A-component does not change significantly. Both show the same rapid condensation behavior at \tc as before. 

Below $x\approx 0.20$ the change in DOS of the B-spin drops to a lower rate, but another phenomenon emerges. It appears visually like a broadening of the superconducting transition, although we know from nuclear relaxation that this is not the case \cite{Lee2026}. This broadening affects both components equally. Perhaps the B-component provides some sort of tuning between the DOS of both components for a coupling to appear.

In other words, as one dopes the cuprates, two metallic components appear, both coupled, but only the isotropic component gains DOS linearly and weakens the pseudogap coupling. This seems consistent with the increase in carrier density \cite{Putzke2021}. The more rapid increase from $p$ to $1+p$, in the region of the phase diagram between near optimal doping to the end of the superconducting dome, correlates with the more rapid change in the $B$-associated DOS in Fig.~\ref{fig:fig1}. 
How the formation of Fermi surface arcs/and or pockets in the underdoped regime \cite{Meng2009} relate to the overall change in DOS  and the pseudogap \cite{Reber2012, Kondo2011} remains to be seen.

\begin{figure}
\centering
\includegraphics[width=0.8\textwidth ]{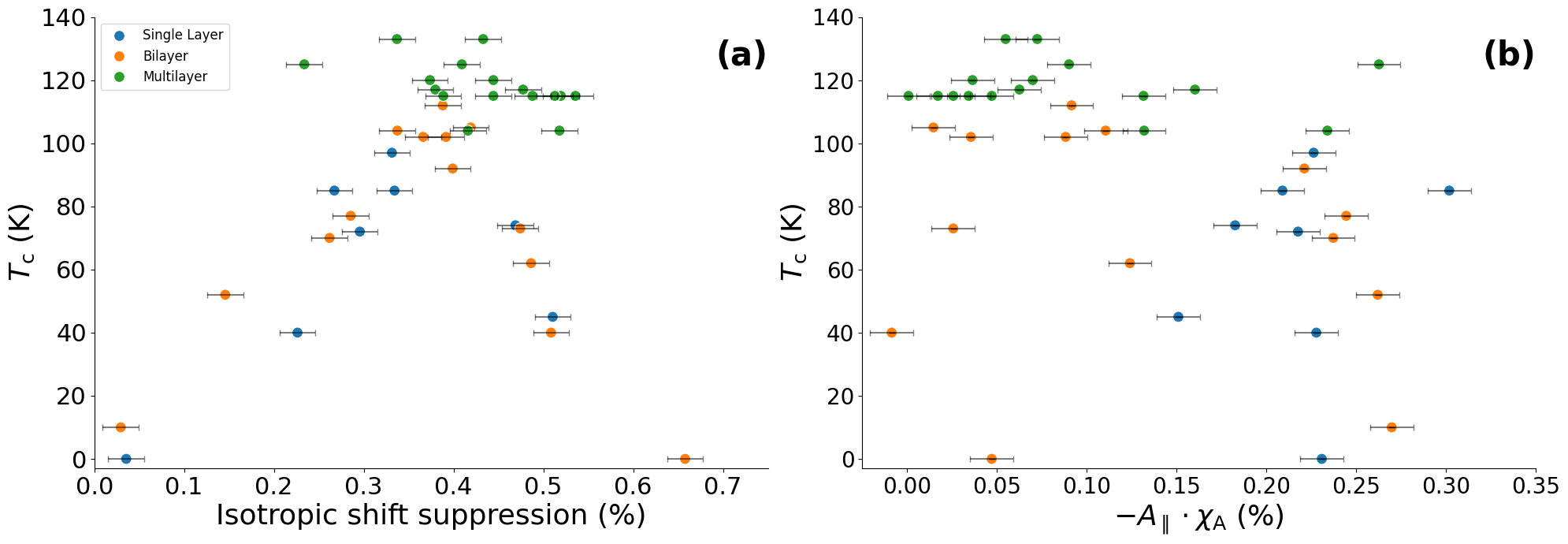}
\caption{High-temperature spin components and $T_\mathrm{c}$. (a) Given the dependence of $T_\mathrm{c}$  and $B\chi_{\mathrm{B}}$ on doping, it is not surprising that the evolution of $B\chi_{\mathrm{B}}$ traces out the superconducting dome. (b) A relationship between \tc and $-A_\parallel\chi_{\mathrm{A}}$ is not readily apparent.}\label{fig:fig5}
\end{figure}

As expected, the high-temperature B-spin draws the superconducting dome when plotted against \tc, as~Fig.~\ref{fig:fig5} shows. The variation in the A-component might be related to family dependent corrugation of the plane \cite{Chmaissem1999,Radaelli1993}.

The size of the pseudogap can be determined, from the changes in the planar O NMR high-temperature shift, as done before \cite{Nachtigal2020,Bandur2026}, as its width in temperature sets the energy scale of the pseudogap. In other words, the pseudogap is apparent in the shifts, but not Cu relaxation, and it involves the broadening of the shifts at \tc.

We know from planar O relaxation that a temperature independent pseudogap is accompanied by the loss in the low-energy states with antiferromagnetic correlations in the A-spin system \cite{Nachtigal2020,Lee2026}. As doping decreases, the gap energy increases accordingly. Specific heat data \cite{Loram1998,Tallon2022} find a similar scenario as planar O relaxation, and is also given by the high-temperature variation of all shifts. 

\begin{figure}
\centering
\includegraphics[width=0.6\textwidth ]{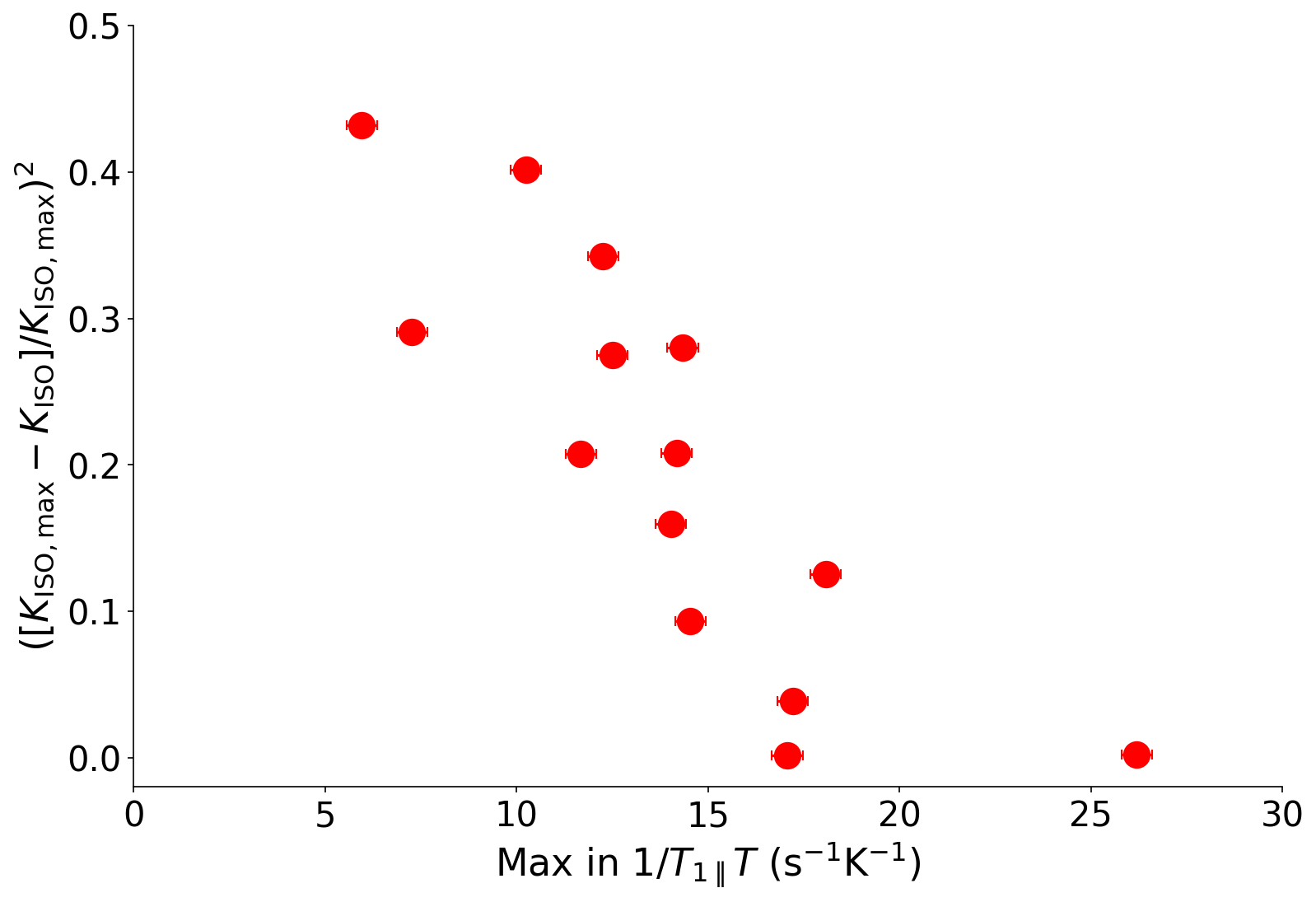}
\caption{Relative isotropic shift suppression squared vs maximum in 1/$T_{1\parallel}T$.  A decrease in isotropic shift suppression (or, conversely, and increase in ${K_{\mathrm{ISO}}}^2$) correlates with an increase in 1/$T_{1\parallel}T$), as both quantities are doping dependent. }\label{fig:fig6}
\end{figure}

We do not know as much detail about the B-spin, except that it changes DOS in the pseudogap region and develops the same broadening at \tc that spreads to higher energies as doping is lowered. The change in DOS is in qualitative agreement with planar Cu relaxation \cite{Lee2026}, that has two metallic components, proportional in their temperature dependence, but only the B-spin is doping dependent, leading to a doping dependent anisotropy. Due to the small $A_\perp$, 1/$T_{1\parallel}T$ is expected to relate primarily to $K_\mathrm{ISO}$. This is shown in Fig.~\ref{fig:fig6}.  \par\medskip

Finally, we discuss the temperature dependence of the shifts, which should hold clues for the condensation behavior. 
Note that the high-temperature shifts can be very different due to doping or material, nevertheless, if we plot both components against each other with temperature as an implicit parameter, we observe fine details, i.e. both shifts decrease by switching between one of three ratios in certain temperature windows, cf.~Fig.~\ref{fig:fig4}.

The first slope \eqref{eq:slope1} is given by $\Delta \chi_\mathrm{B}\neq 0$ and $\Delta \chi_\mathrm{A} =0$, i.e.\@ only the B-spin condenses (as it can be very large due to doping). We observe such behavior for the high-temperature plateaus as a function of doping that changes $\chi_\mathrm{B}$ DOS only. It is also observed as function of temperature (it is the reason why 'isotropic shift lines' dominate the ${^{63}K}_\perp(T)$ vs ${^{63}K}_\parallel(T)$ plot \cite{Haase2017}). When the smaller A-spin has already vanished and the B-spin is not fully condensed, we observe this slope. Note that we do not observe the opposite, i.e. a changing A-spin while the B-spin remains constant. 

The other two slopes appear only as function of temperature (not doping) and document simultaneous changes of both components. Slope \eqref{eq:slope2} is given by $|A_\parallel|\Delta_T\chi_\mathrm{A}(T) = B \Delta_T\chi_\mathrm{B}(T)$. Slope \eqref{eq:slope3} shows half the A-spin change per B-spin compared to \eqref{eq:slope2}. One wonders if these different condensation behaviors could be related to the widely-discussed gap symmetry admixture \cite{Keller2008, BussmannHolder2024, Khasanov2007b}.


Now, we focus on overdoped, superconducting materials (no pseudogap) for which condensation occurs at \tc (sharp drop). Rather typical is the initial drop with \eqref{eq:slope2}, i.e. A-spin and B-spin condensing at the same rate. Once A-spin is fully condensed (total loss of NMR shift component), the initially rather large B-spin continues to drop until it is also fully condensed. A few materials, e.g. \ybcoF and optimally doped \ybco, after an initial drop, switch to slope \eqref{eq:slope3}, which we discuss now.

Materials with a pseudogap, i.e.\@ with broadening at \tc, are special since the shifts already become temperature dependent above \tc. Interestingly, most materials assume immediately slope \eqref{eq:slope3} and follow this slope until one of the components is fully condensed (underdoped or overdoped) or both disappear together (optimally doped). In other words, there is no change of slope at \tc. In such materials, the B-spin condenses twice as fast, which points to its different origin. 
Optimally-doped and higher-\tc materials follow the slope \eqref{eq:slope3} for the whole temperature range, such that both Cu shifts and the planar O shift are all proportional to each other and match as a function of temperature. 
A natural explanation for this behavior is that spin pairing occurs when the shifts become temperature dependent, i.e. from the high-temperature metal. At \tc, phase coherence among the pairs sets in and nuclear relaxation drops.

Our results appear not to favor a single-band picture with two spin components that act very differently with doping (while the condensation scenarios might effectively be viewed through simpler models). In a multi-band description, the Cu orbital shift anisotropy may be closer to its ionic value, ${^{63}K}_{\mathrm{L}\parallel} \sim 4\; {^{63}K}_\mathrm{L\perp}$ \cite{Pennington1989}. This would solve the orbital shift problem, for otherwise one has to assume an additional, largely material and temperature independent negative A-spin to account for the observed low-temperature shift offset for ${^{63}K}_{\parallel}$. 

While we make essentially no assumptions when analyzing the temperature dependent shifts of the cuprates and discover features that relate to the cuprate phase diagram, we find no clear indicator for the maximum \tc of a family, $T_\mathrm{c,max}$. This information does not seem to be directly encoded in the magnetic shifts of the cuprates. A very clear correlation, however, was found with Cu nuclear relaxation anisotropy \cite{Lee2026}. It is related to B-spin DOS, but that obviously does not capture the full dependence. In any case, the magnetic relaxation does relate to $T_\mathrm{c,max}$. It thus also correlates with the planar O hole content ($2n_\mathrm{p}$) at optimal doping, which is nearly proportional to the maximum \tc of the cuprates ($T_\mathrm{c,max} \approx 200 \text{ K } 2n_\mathrm{p}$ \cite{Rybicki2016}). The latter relation is confirmed by DMFT \cite{Kowalski2021}, which goes towards understanding the cuprates by first-principles \cite{Bacq2025}. 

We also note that we did not make any assumptions about the transferred spin density for the B-component. In a single band picture, $B=4B'$ was assumed as the transfer originates in the 4 Cu neighbors \cite{Mila1989b}. The change of the B-spin across the phase diagram does not indicate changes in the transfer, but the different condensation slopes might be connected with it, as \lsco is rather different in terms of condensation and relaxation \cite{Jurkutat2019}. Stripes \cite{Tranquada1995} might alter the transfer of spin as well. The neutron spin response should be tied to the A-spins, but the presence of the B-spin should also influence the response at the antiferromagnetic wave vector.

It will be interesting to connect our robust NMR phenomenology of shifts and relaxation \cite{Lee2026} to other probes' findings in greater detail.

\begin{acknowledgements}
We acknowledge financial support from Leipzig University, in particular Roger Gläser, and stimulating discussions with Nabeel Aslam. 
\end{acknowledgements}

{\flushleft \bf Author contributions\\}
A.L. performed all data analysis. A.L. and J.H. contributed equally to preparing the manuscript. J.H. had the overall leadership.

{\flushleft \bf Competing Interests\\}
{ There are no competing interests for any of the authors.}\par\medskip

{\flushleft \bf Data Availability Statement \\}
All used data will be made available which enables reproduction of the results.\par\medskip

{\flushleft \bf Funding\\}
Funding of the research came from Leipzig University.\par\medskip
\vspace{0.5cm}

\bibliography{JH-Cuprate.bib}   

\printindex
 
\newpage
\appendix
\section{Decomposition of the two spin components}\label{sec:decomposition}
For the analysis, we need to set up a basic description of the shifts. As mentioned in the main text, the total Cu and O shifts fail to scale with each other as a function of temperature as one would expect for a single susceptibility. In fact, even the Cu shifts measured with the applied field parallel and perpendicular to the CuO$_2$ planes fail to scale. The reason for this failure has recently been shown by some of us to be due to the absence at planar O of the additional isotropic shift seen on planar Cu \cite{Bandur2026}. This is demonstrated by the fact that, surprisingly, the planar O shift is set by the Cu axial shift, i.e., 
\begin{equation}
{^{63}K}_{\rm axial} (T,x) = {^{63}K}_\perp (T,x) - {^{63}K}_\parallel (T,x) \propto  {^{17}K}_{\rm c} (T,x) +\delta,\label{eq:s1a}
\end{equation}
where $\delta$ is an experimentally determined temperature independent offset \cite{Bandur2026}. 

In order to describe these findings in greater detail, we introduce the following equations for the doping ($x$) and temperature (T) dependent planar Cu and O spin shifts: 
\begin{align}
{^{63}K}_\perp (T,x) &= B \chi_{\rm B}(T,x) + A_\perp \chi_{\rm A}(T,x) \approx B \chi_{\rm B}(T,x) \label{eq:s2a}\\
{^{63}K}_\parallel (T,x) &= B \chi_{\rm B}(T,x) + A_\parallel \chi_{\rm A}(T,x) + \Delta_\parallel(x),\label{eq:s3a}\\
{^{17}K}_{\rm c} (T,x) &= C_{\rm c} \chi_{\rm A}(T,x),\label{eq:s4a}
\end{align}
wherein $\chi_{\rm A}$ and $\chi_{\rm B}$ may include a coupling term. 
Since the Cu orbital shift for \cpara is not reliable, we have introduced another variable $\Delta_\parallel(x)$ which is the difference to the actual orbital shift not known to us. The onsite hyperfine constant, $A_\alpha$, relating to the $3d(x^2-y^2)$ electrons, is known to be rather anisotropic and negative for \cpara ($|A_\parallel |/ A_\perp \approx 6$,\cite{Pennington1989,Huesser2000}), while the transferred spin enters isotropically via the hyperfine $B$ and thus changes both shifts identically. Thus with eqns. \eqref{eq:s2a} to \eqref{eq:s4a}, we should be able to describe all cuprate shifts and learn about the susceptibilities as a function of doping and temperature.

In this analysis, we utilize the aforementioned anisotropy of the one term to obtain the contributions to the shifts from both the isotropic and anisotropic terms individually without needing to know the precise values of the hyperfine constants. Furthermore, we show how these contributions evolve as a function of doping and family. In order to quantify the distinct isotropic and anisotropic terms, one needs to assume a reasonable reference. It is well known that while ${^{63}K}_\perp (T\rightarrow 0) \approx 0.35\%$, i.e., to very similar low-temperature point, which agrees with the expected orbital shift from first principle calculations \cite{Renold2003}. This is not the case for ${^{63}K}_\parallel$ since the low-temperature shifts for this direction are spread out and appear to carry a significant family dependence. First principle calculations suggest an orbital term of about 0.72\% \cite{Renold2003} due to the hybridization with planar O. Since we cannot solve this ambiguity here, we leave this orbital term undefined currently, and reference instead to the low temperature values of ${^{63}K}_\alpha$. For a few materials, the low temperature shifts are not available. Since materials within a family tend to end at the same low temperature shift values (see, for example, Hg1201), members of the same family for which the shifts were measured down to the lowest temperatures were used to estimate the low temperature shift those for which the shifts were only measured at relatively higher temperatures.

Making use of such a reference point, and the anisotropy of the $A$ hyperfine, we can obtain the iso- and anisotropic contributions to any given shift using
\begin{equation}
^{63}K_\parallel-^{63}K_\parallel(T\rightarrow0) = -K_{ANI}+K_{ISO}\label{eq:kpar}
\end{equation}
and 
\begin{equation}
^{63}K_\perp-^{63}K_\perp(T\rightarrow0)=\frac{1}{6}K_{ANI}+K_{ISO}\label{eq:kperp}
\end{equation}

In order to obtain the suppression of the isotropic shift component within this framework, we compare the calculated contributions to that of the most highly overdoped materials for which the required data is available. Note that, although some suppression could still be present in these materials (presumably small considering the isotropic relaxation \cite{Jurkutat2019}), the overall phenomenology of the doping dependence remains the same. Additionally, we have verified that small changes in the anisotropy of the hyperfine $A$ have little effect on the overall phenomenology.


\section{Literature Data}
The table below contains the references for all shift and relaxation data used. The values listed in the doping column for single and bilayer materials are those one finds if estimating from the parabolic dependence of \tc on doping alone, while those listed for multilayer materials are taken as given from the reference where available. UD, OP, and OD in the material labels stand for underdoped, optimally doped, and overdoped respectively and are followed by the quoted value of \tc. Materials with distinct inner and outer CuO$_2$ planes are labeled additionally with -IP and -OP to signify data pertaining to the inner and outer planes respectively.\\
\begin{table}[h]
\tiny
\centering
\begin{tabular}{|c|c|c|c|c|c|}
\hline
Material label & Formula & Doping& $^{63}K_\parallel$ & $^{63}K_\perp$ & 1/$^{63}T_{1\parallel}T$ \\
\hline
Hg1201UD45K & HgBa$_2$CuO$_{4+\delta}$ & 0.06 & \cite{Rybicki2015} & \cite{Rybicki2015} & -  \\
\hline
Hg1201UD74K & HgBa$_2$CuO$_{4+\delta}$ & 0.1 & \cite{Rybicki2015} & \cite{Rybicki2015} & \cite{Gippius1999} \\
\hline
Hg1201OP97K & HgBa$_2$CuO$_{4+\delta}$ & 0.16 & \cite{Rybicki2015} & \cite{Rybicki2015} & -  \\
\hline
Hg1201OD85K & HgBa$_2$CuO$_{4+\delta}$ & 0.19 & \cite{Rybicki2015} & \cite{Rybicki2015} & \cite{Gippius1999} \\
\hline
Tl2201OP85K & Tl$_2$Ba$_2$CuO$_{6+y}$ & 0.16 & \cite{Kambe1993} & \cite{Kambe1993} & - \\
\hline
Tl2201OD72K & Tl$_2$Ba$_2$CuO$_{6+y}$ & 0.2 & \cite{Fujiwara1991} & \cite{Fujiwara1991} & \cite{Fujiwara1991}\\
\hline
Tl2201OD40K & Tl$_2$Ba$_2$CuO$_{6+y}$ & 0.24 & \cite{Fujiwara1991} & \cite{Fujiwara1991} & \cite{Fujiwara1991}\\
\hline
Tl2201OD0K & Tl$_2$Ba$_2$CuO$_{6+y}$ & 0.27 & \cite{Fujiwara1991}& \cite{Fujiwara1991} & \cite{Fujiwara1991} \\
\hline
Tl1212OD70K & TlSr$_2$CaCu$_2$O$_{7-\delta}$ & 0.21 & \cite{Magishi1996} & \cite{Magishi1996} & \cite{Magishi1996} \\
\hline
Tl1212OD52K & TlSr$_2$CaCu$_2$O$_{7-\delta}$ & 0.23 & \cite{Magishi1996}  & \cite{Magishi1996}  & \cite{Magishi1996}\\
\hline
Tl1212OD10K & TlSr$_2$CaCu$_2$O$_{7-\delta}$ & 0.265 & \cite{Magishi1996}  & \cite{Magishi1996}  & \cite{Magishi1996}\\
\hline
YBCO6 & YBa$_2$Cu$_3$O$_6$ & 0 & 
\cite{Mali1991} & \cite{Mali1991} & - \\
\hline
YBCOUN62 & YBa$_2$Cu$_3$O$_{6.63}$ & 0.11 & \cite{Takigawa1991} & \cite{Takigawa1991} & \cite{Takigawa1991} \\
\hline
YBCO7 & YBa$_2$Cu$_3$O$_7$ & 0.17 & \cite{Takigawa1989} & \cite{Takigawa1989} & \cite{Walstedt1989}
 \\
\hline
Y1248 & YBa$_2$Cu$_4$O$_8$ & 0.12 & \cite{Bankay1994} & \cite{Bankay1994} & \cite{Zimmermann1991} \\
\hline
Tl2212UN102K & Tl$_2$Ba$_2$CaCu$_2$O$_{8-\delta}$ & 0.127 & \cite{Gerashchenko1999} & \cite{Gerashchenko1999} & \cite{Gerashchenko1999}  \\
\hline
Tl2212OP112K & Tl$_2$Ba$_2$CaCu$_2$O$_{8-\delta}$ & 0.16 & \cite{Gerashchenko1999} & \cite{Gerashchenko1999} & \cite{Gerashchenko1999} \\
\hline
Tl2212OD104K & Tl$_2$Ba$_2$CaCu$_2$O$_{8-\delta}$ & 0.189 & \cite{Gerashchenko1999} & \cite{Gerashchenko1999} & \cite{Gerashchenko1999}  \\
\hline
Ba0212UN40K & Ba$_2$CaCu$_2$O$_6$(F,O)$_2$ & 0.085 & \cite{Shimizu2011} & \cite{Shimizu2011} & - \\
\hline
Ba0212UN73K & Ba$_2$CaCu$_2$O$_6$(F,O)$_2$ & 0.114 & \cite{Shimizu2011} & \cite{Shimizu2011} & - \\
\hline
Ba0212OP105K & Ba$_2$CaCu$_2$O$_6$(F,O)$_2$ & 0.149 & \cite{Shimizu2011} & \cite{Shimizu2011} & -\\
\hline
Ba0212OD102K & Ba$_2$CaCu$_2$O$_6$(F,O)$_2$ & 0.174 & \cite{Shimizu2011} & \cite{Shimizu2011} & -\\
\hline

Bi2212OD77K & Bi$_2$Sr$_2$CaCu$_2$O$_{8+\delta}$ & 0.2 & \cite{Walstedt1991} & \cite{Walstedt1991} & - \\
\hline
Ba0223OP120K-IP & Ba$_2$Ca$_2$Cu$_3$O$_6$(F,O)$_2$ & 0.16 & \cite{Shimizu2011} & \cite{Shimizu2011} & - \\
\hline
Ba0223OP120K-OP & Ba$_2$Ca$_2$Cu$_3$O$_6$(F,O)$_2$ & 0.16 & \cite{Shimizu2011} & \cite{Shimizu2011} & - \\
\hline
Tl2223UD104K-IP & Tl$_2$Ba$_2$Ca$_2$Cu$_3$O$_{10-\delta}$ & 0.115 & \cite{Piskunov1998} & \cite{Piskunov1998}&- \\
\hline
Tl2223UD104K-OP & Tl$_2$Ba$_2$Ca$_2$Cu$_3$O$_{10-\delta}$ & 0.115 & \cite{Piskunov1998}& \cite{Piskunov1998} & - \\
\hline
Tl2223OP125K-IP & Tl$_2$Ba$_2$Ca$_2$Cu$_3$O$_{10-\delta}$ & 0.16 & \cite{Han1994} & \cite{Han1994} & -\\
\hline
Tl2223OP125K-OP & Tl$_2$Ba$_2$Ca$_2$Cu$_3$O$_{10-\delta}$ & 0.16 & \cite{Han1994} & \cite{Han1994}& -\\
\hline
Tl2223OD117K-IP & Tl$_2$Ba$_2$Ca$_2$Cu$_3$O$_{10-\delta}$ & 0.156 & \cite{Piskunov1998} & \cite{Piskunov1998} & - \\
\hline
Tl2223OD117K-OP & Tl$_2$Ba$_2$Ca$_2$Cu$_3$O$_{10-\delta}$ & 0.21 & \cite{Piskunov1998} & \cite{Piskunov1998} & -\\
\hline
Tl2223OD115K-IP & Tl$_2$Ba$_2$Ca$_2$Cu$_3$O$_{10-\delta}$ & 0.124 & \cite{Zheng1996} & \cite{Zheng1996} &  \cite{Zheng1996} \\
\hline
Tl2223OD115K-OP & Tl$_2$Ba$_2$Ca$_2$Cu$_3$O$_{10-\delta}$ & 0.196 & \cite{Zheng1996} & \cite{Zheng1996} &  \cite{Zheng1996} \\
\hline
Hg1223UD115K-IP & HgBa$_2$Ca$_2$Cu$_3$O$_{8-\delta}$ & 0.053 & \cite{Julien1996} & \cite{Julien1996} &  - \\
\hline
Hg1223UD115K-OP & HgBa$_2$Ca$_2$Cu$_3$O$_{8-\delta}$ & 0.177 & \cite{Julien1996} & \cite{Julien1996} &  -\\
\hline
Hg1223OP133K-IP & HgBa$_2$Ca$_2$Cu$_3$O$_{8-\delta}$ & 0.14 & \cite{Magishi1995} & \cite{Magishi1995} &  \cite{Magishi1995} \\
\hline
Hg1223OP133K-OP & HgBa$_2$Ca$_2$Cu$_3$O$_{8-\delta}$ & 0.19 & \cite{Magishi1995} & \cite{Magishi1995} &  \cite{Magishi1995} \\
\hline
Hg1245OP115K-IP & HgBa$_2$Ca$_4$Cu$_5$O$_y$ & 0.1175 & \cite{Kotegawa2004} & \cite{Kotegawa2004} & - \\
\hline
Hg1245OP115K-OP & HgBa$_2$Ca$_4$Cu$_5$O$_y$ & 0.2025 & \cite{Kotegawa2004} & \cite{Kotegawa2004} & - \\
\hline
\end{tabular}
\end{table}
\end{document}